\documentclass[useAMS,usenatbib]{mn2e}
\usepackage{graphicx,txfonts}

\def\ch2{$\chi^2$}

\newcommand{\MOLH}{\hbox{${\rm H}_2$}}
\newcommand{\kms}{\hbox{${\rm km\,s}^{-1}$}}
\newcommand{\zem}{\hbox{$z_{\rm em}$}}
\newcommand{\zabs}{\hbox{$z_{\rm abs}$}}
\newcommand{\HI}{\hbox{{\rm H}{\sc \,i}}}
\newcommand{\NHI}{\hbox{$N(\HI)$}}
\newcommand{\scm}{\hbox{${\rm cm}^{-2}$}}
\newcommand{\EW}{\hbox{$W_{\rm r}^{\lambda2796}$}}
\newcommand{\hkpc}{\hbox{$h^{-1}$~kpc}}

\defcitealias{LedouxC_06a}{L06}
\defcitealias{BoucheN_06a}{B06}

\def\lapp{\ifmmode\stackrel{<}{_{\sim}}\else$\stackrel{<}{_{\sim}}$\fi}
\def\gapp{\ifmmode\stackrel{>}{_{\sim}}\else$\stackrel{>}{_{\sim}}$\fi}

\def\bsp_small{\vspace{0.5cm}\small\noindent This paper has been typeset
from a \TeX/\LaTeX\ file prepared by the author.\normalsize}

\title[Metallicity and kinematics of (sub-)DLAs]{The connection
  between metallicity and metal-line kinematics in (sub-)damped
  Lyman-{\boldmath $\alpha$} systems} \author[Murphy et al.]
  {M.~T.~Murphy$^{1}$\thanks{E-mail: mim@ast.cam.ac.uk},
  S.~J.~Curran$^{2}$, J.~K.~Webb$^{2}$, H.~M\'enager$^{1,2}$, B.~J.~Zych$^{1}$\\
  $^{1}$Institute of Astronomy, University of Cambridge, Madingley Road,
  Cambridge CB3 0HA\\
  $^{2}$School of Physics, University of New South Wales, Sydney
  NSW 2052, Australia}

\begin{document}

\date{Accepted 2006 December 29. Received 2006 December 29; in
  original form 2006 October 27}

\pagerange{\pageref{firstpage}--\pageref{lastpage}} \pubyear{2006}

\maketitle

\label{firstpage}

\begin{abstract}
  A correlation between the metallicity, [M/H], and rest-frame Mg{\sc
    \,ii} equivalent width, \EW, is found from 49 DLAs and strong
  sub-DLAs drawn from the literature over the redshift range
  $0.2<\zabs<2.6$.  The correlation is significant at $4.2\,\sigma$
  and improves to $4.7\,\sigma$ when the mild evolution of [M/H] with
  redshift is taken into account. Even when including only the 26 DLAs
  (i.e.~exlcuding sub-DLAs) which have Zn-metallicities and
  $\EW>0.7$\,\AA, the correlation remains at $>3\,\sigma$
  significance. Since the Mg{\sc \,ii} $\lambda$2796 transition is
  predominantly saturated in DLAs (which always have \EW\ greater than
  0.3\,\AA), \EW\ is far more sensitive to the kinematic spread of the
  metal velocity components across the absorption profile than it is
  to [M/H]. Thus, the observed [M/H]--\EW\ correlation points to a
  strong link between the absorber metallicity and the mechanism for
  producing and dispersing the velocity components. We also note that
  approximately half of the 13 known \MOLH\ absorbers have very high
  \EW\ and very broad velocity structures which show characteristics
  usually associated with outflows.  Follow-up ultraviolet- and
  blue-sensitive high-resolution spectra of high-\EW\ systems,
  initially identified in low-resolution spectra, may therefore yield
  a large number of new \MOLH\ discoveries.
\end{abstract}


\section{Introduction}\label{sec:intro}

The damped Lyman-$\alpha$ absorbers (DLAs) observed along sight-lines
to quasars (QSOs) are potentially important pieces in the puzzle of
galaxy formation. They are the highest column density absorption
systems, with $\NHI \ge 2\times10^{20}{\rm \,cm}^{-2}$, and the first
DLA surveys quickly established that they contain a large fraction of
the high-$z$ baryons {\it available} for star formation
\citep{WolfeA_86a,LanzettaK_91a}. The early estimates of the neutral
gas mass density at absorption redshifts $\zabs\sim3$ were similar to
the estimates of the mass density in stars at $z=0$. It was therefore
natural to assume that DLAs act as the neutral gas reservoirs for star
formation. However, firmly establishing the link between DLAs and star
formation relies on our understanding of the absorber--galaxy
connection. Direct imaging of DLAs (or strong Mg{\sc \,ii} systems;
see below) at $\zabs<1.5$ reveals the hosts to be a mix of irregulars,
dwarfs, spirals and low surface-brightness galaxies
\citep[e.g.][]{BergeronJ_91a,LeBrunV_97a,RaoS_03a,ChenH-W_03a}. This
is further borne out by a blind 21-cm emission survey at $z=0$
\citep{Ryan-WeberE_03a}. However, quantifying the mix of DLA
host-galaxy morphologies from low- to high-redshift is fraught with
selection effects, not least of which are luminosity bias against
faint galaxies and the proximity of some host-galaxies to the QSO
sight-line. Indeed, very few DLA host-galaxies have been identified at
$\zabs\ga2$
\citep[e.g.][]{MollerP_93a,DjorgovskiS_96a,ChristensenL_05a,WeatherleyS_05a}.

The Mg{\sc \,ii} $\lambda$2796/2803 doublet has proved very important
for exploring the absorber--galaxy connection at $0.3\la\zabs\la2.6$
where it is observable with optical QSO absorption spectroscopy.
Indeed, many of the $\zabs<1.5$ imaging studies above initially
identified absorption systems via the easily recognised Mg{\sc \,ii}
doublet \citep[see also,
e.g.,][]{LanzettaK_90a,BergeronJ_92a,SteidelC_92a,DrinkwaterM_93a,SteidelC_94a}.
Also, since the ionization potential of Mg{\sc \,i} is $<13.6{\rm
  \,eV}$ but that of Mg{\sc \,ii} is $>13.6{\rm \,eV}$, the latter
traces cold gas.  Attempts have therefore been made to identify DLA
candidates at low-\zabs\ via their Mg{\sc \,ii} absorption
\citep[e.g.][]{RaoS_95a,RaoS_06a,EllisonS_06a}. The Mg{\sc \,ii}
doublet transitions are also among the strongest metal lines observed
in QSO absorbers, meaning that (i) the Mg{\sc \,ii} $\lambda$2796
rest-frame equivalent width in DLAs is typically $\EW\ga0.3$\,\AA,
allowing such systems to be identified in relatively low-resolution
spectra, and (ii) most velocity components across the Mg{\sc \,ii}
profile are saturated in DLAs. This implies that \EW\ is mainly
sensitive to the kinematic extent of the velocity components, $\Delta
v$, rather than their column density. A good demonstration of this is
figure 3 of \citet{EllisonS_06a}.

In this paper we search for a relationship between \EW\ and the
metallicity\footnote{Metallicity is defined as the heavy element
  abundance with respect to hydrogen, relative to that of the solar
  neighbourhood: ${\rm [M/H]} \equiv \log[N({\rm M})/N({\rm H})] -
  \log[N({\rm M})/N({\rm H})]_\odot$}, [M/H], of DLAs and strong
sub-DLAs. Since \EW\ is a measure of the kinematic spread of the cold,
metal-bearing velocity components, such a relationship may constrain
the mechanisms for producing and dispersing the metals. This may guide
our understanding of the physical nature of DLAs.

There currently exist only indirect and possibly contradictory
constraints on a [M/H]--\EW\ relationship for DLAs and sub-DLAs.
\citet{NestorD_03a} suggest that [M/H] increases with increasing \EW.
They identified strong Mg{\sc \,ii} systems in Sloan Digital Sky
Survey (SDSS) QSO spectra over the redshift range $0.9<\zabs<2.0$ and
constructed composite spectra in the absorber rest-frame. The Zn{\sc
  \,ii} lines were stronger in the composite of $\EW\ge1.3$\,\AA\
absorbers compared with those in the composite of $1.0<\EW<1.3$\,\AA\
systems. Assuming that the mean \NHI\ is not significantly different
in these two regimes, they conclude that some evidence exists for a
[Zn/H]--\EW\ correlation. On the other hand, \citet{YorkD_06a} find
less direct evidence for an anti-correlation between [Zn/H] and \EW.
From a series of composite SDSS spectra with various mean \EW\ they
find that the dust-reddening, $E(B-V)$, caused by the absorbers rises
sharply with increasing \EW. While the Zn{\sc \,ii} line-strengths
also increase, they show a slower increase than $E(B-V)$ over the same
range of \EW. The composite spectra also suggest a dust extinction
curve typical of the Small Magellanic Cloud (SMC), leading
\citeauthor{YorkD_06a} to assume a constant dust-to-gas ratio,
i.e.~$\NHI\propto E(B-V)$, as observed in the SMC. Thus, since the
Zn{\sc \,ii} column density increases slower than the \NHI\ inferred
from $E(B-V)$ under this assumption, \citet{YorkD_06a} argue that
[Zn/H] decreases with increasing \EW. We discuss these results further
in light of our new results in Section \ref{sec:disc}.

This paper is organised as follows. Section \ref{sec:corr} explains
our sample selection and explores an evident [M/H]--\EW\ relationship
in that sample. Section \ref{sec:disc} compares our results with the
[M/H]--$\Delta v$ correlation recently discovered by
\citet{LedouxC_06a} at (generally) higher redshifts. We also discuss
whether our observed [M/H]--\EW\ relationship can be taken as evidence
for a correlation or anti-correlation between absorber metallicity and
host-galaxy mass.  Finally, we discuss whether a large fraction of
strong Mg{\sc \,ii} absorbers might arise in outflows from relatively
low mass galaxies.  In particular, about half of those absorbers in
which H$_2$ has been detected show some evidence of an outflow origin.

\section{Correlation between metallicity and Mg{\sc \,ii}
  equivalent width in DLAs and sub-DLAs}\label{sec:corr}

\subsection{Sample definition}

\begin{table*}
\begin{center}
\vspace{-1mm}
\begin{minipage}{0.905\textwidth}
  \caption{DLAs and sub-DLAs where both Mg{\sc \,ii} $\lambda$2796
    \AA\ equivalent width, \EW, and metallicity, [M/H], have been
    measured. B1950 QSO names are given. \NHI\ is the total neutral
    hydrogen column density [\scm], \zabs\ is the DLA redshift and
    \zem\ is the redshift of the background QSO. Errors on \EW\ and
    [M/H] are quoted from the literature but in many cases do not
    include continuum-fitting uncertainties. [M/H] is measured
    relative to the solar abundances of \citet{LoddersK_03a} in most
    cases (see text). The element used to determine [M/H] is Zn where
    possible; when this is not available we corrected for dust
    depletion via ${\rm [M/H]} = {\rm [Fe/H]} + 0.4 = {\rm [Cr/H]} +
    0.2$ \citep[see][]{ProchaskaJ_03b}.}
\label{dla}
\vspace{-3mm}
\begin{tabular}{@{}l c c c c c  c c c l}\hline
QSO         & $\log\NHI$ & \zabs   & \zem    & \EW\ [\AA]      & Ref.               & [M/H]           & M  & Ref.     & Notes        \\\hline
0013$-$0029 & 20.8       & 1.973   & 2.08694 & $4.7\pm0.3    $ & R03                & $-0.72\pm0.09$  & Zn & P02      & $a$          \\
0058$+$0155 & 20.1       & 0.61251 & 1.954   & $1.68\pm0.03  $ & HIRES              & $+0.05\pm0.21$  & Zn & P00      & ---          \\
0100$+$130  & 21.4       & 2.3091  & 2.681   & $0.85\pm0.01  $ & UVES$^{7,19}$      & $-1.54\pm0.09$  & Zn & PW99     & PHL 957      \\
0235$+$164  & 21.7       & 0.52385 & 0.94    & $2.34\pm0.05  $ & LB92               & $-0.22\pm0.15$  & Zn & J04,P03a & $b$          \\
0302$-$223  & 20.4       & 1.00945 & 1.409   & $1.16\pm0.04  $ & R06                & $-0.56\pm0.12$  & Zn & P00      & $e$          \\
0405$-$4418 & 21.0       & 2.5505  & 3.00    & $2.38\pm0.25  $ & UVES$^{10,11,14}$  & $-1.32\pm0.11$  & Zn & LE03     & CTQ 0247     \\
0405$-$4418 & 21.1       & 2.595   & 3.00    & $1.29\pm0.09  $ & UVES$^{10,11,14}$  & $-1.04\pm0.10$  & Zn & LE03     & $a$, CTQ 0247\\
0454$+$039  & 20.7       & 0.8596  & 1.345   & $1.45\pm0.01  $ & C00                & $-0.99\pm0.12$  & Zn & P00      & $c$          \\
0458$-$020  & 21.7       & 2.03945 & 2.286   & $1.69\pm0.07  $ & W93                & $-1.19\pm0.09$  & Zn & PW99     & $b$          \\
0512$-$3329 & 20.5       & 0.931   & 1.569   & $1.94         $ & E06b               & $-1.09\pm0.11$  & Fe & L05      & ---          \\
0515$-$4414 & 20.5       & 1.1508  & 1.713   & $2.34\pm0.02  $ & UVES$^{1,5,16}$    & $-0.95\pm0.19$  & Zn & D00      & $a$          \\
0528$-$250  & 20.6       & 2.141   & 2.813   & $1.89\pm0.08  $ & UVES$^{6,9,11}$    & $-1.45\pm0.07$  & Zn & C03a     & $d$          \\
0551$-$366  & 20.4       & 1.962   & 2.318   & $5.80\pm0.08  $ & R03                & $-0.11\pm0.09$  & Zn & L02b     & $a$          \\
0738$+$313  & 20.8       & 0.2212  & 0.635   & $0.61\pm0.04  $ & B87                & $-1.24\pm0.2 $  & Cr & K04      & $b$          \\
0827$+$243  & 20.3       & 0.5247  & 0.939   & $2.90         $ & RT00               & $-0.62\pm0.08$  & Fe & K04      & $b$          \\
0841$+$1256 & 21.3       & 2.3745  & 2.50    & $0.76\pm0.07  $ & UVES$^{2,15}$      & $-1.47\pm0.15$  & Zn & PW99     & ---          \\
0933$+$732  & 21.6       & 1.4789  & 2.528   & $0.95\pm0.08  $ & R06                & $-1.58\pm0.25$  & Zn & K04      & $e$          \\
0935$+$417  & 20.3       & 1.3726  & 1.98    & $1.04         $ & P04                & $-0.78\pm0.2 $  & Zn & M95      & ---          \\
0945$+$436  & 21.5       & 1.223   & 1.89149 & $1.12\pm0.05  $ & SDSS               & $-0.95\pm0.10$  & Zn & P03a     & ---          \\
0949$+$527  & 20.1       & 1.7678  & 1.87476 & $3.88\pm0.05  $ & SDSS               & $+0.02\pm0.10$  & Zn & P06b     & ---          \\
0952$+$179  & 21.3       & 0.2378  & 1.472   & $0.63\pm0.11  $ & RT00               & $-1.45\pm0.2 $  & Cr & K05      & $b$          \\
0957$+$561A & 20.3       & 1.391   & 1.413   & $2.25         $ & E06b               & $-0.91       $  & Fe & C03b     & $c$          \\
0957$+$561B & 19.9       & 1.391   & 1.413   & $2.13         $ & E06b               & $-0.63       $  & Fe & C03b     & $c$          \\
1101$-$264  & 19.5       & 1.838   & 2.145   & $0.94         $ & E06b               & $-1.06\pm0.07$  & Fe & D03      & ---          \\
1104$-$1805 & 20.9       & 1.6614  & 2.319   & $0.96         $ & L99                & $-0.99\pm0.02$  & Zn & L99      & ---          \\
1104$+$0104 & 21.0       & 0.7404  & 1.3924  & $2.95\pm0.03  $ & R06                & $-0.60\pm0.2 $  & Zn & K04      & ---          \\
1122$-$1649 & 20.5       & 0.6819  & 2.40    & $1.72         $ & P04                & $-1.00\pm0.15$  & Fe & L02a,P03a& $d$          \\
1151$+$068  & 21.3       & 1.7736  & 2.762   & $0.52\pm0.05  $ & R03                & $-1.53\pm0.14$  & Zn & P97      & ---          \\
1157$+$014  & 21.8       & 1.94362 & 1.986   & $1.38\pm0.06  $ & R03                & $-1.36\pm0.12$  & Zn & L03      & $b$          \\
1209$+$093  & 21.4       & 2.5840  & 3.297   & $3.20\pm0.50  $ & UVES$^{8,18}$      & $-1.04\pm0.11$  & Zn & P03b     & $e$          \\
1210$+$1731 & 20.6       & 1.892   & 2.543   & $1.09         $ & E06b               & $-0.86\pm0.10$  & Zn & D06      & ---          \\
1215$+$333  & 21.0       & 1.999   & 2.606   & $1.03\pm0.15  $ & SS92               & $-1.25\pm0.09$  & Zn & P99      & $d$          \\
1223$+$1753 & 21.5       & 2.4658  & 2.936   & $1.54\pm0.06  $ & UVES$^{3,4,13}$    & $-1.58\pm0.10$  & Fe & P01      & $e$          \\
1229$-$0207 & 20.8       & 0.39498 & 1.045   & $2.17\pm0.07  $ & LB92               & $-0.45\pm0.19$  & Zn & B98      & $b$          \\
1247$+$267  & 19.9       & 1.22319 & 2.038   & $0.42         $ & E06b               & $-1.02\pm0.25$  & Zn & P99      & $d$          \\
1320$-$0006 & 20.2       & 0.716   & 1.38839 & $2.13\pm0.05  $ & SDSS               & $+0.61\pm0.20$  & Zn & P06a     & ---          \\
1328$+$307  & 21.3       & 0.692154& 0.849   & $0.33         $ & SM79               & $-1.20       $  & Zn & MY92     & $b$, 3C 286  \\
1331$+$170  & 21.2       & 1.77642 & 2.084   & $1.33\pm0.09  $ & R03                & $-1.26\pm0.10$  & Fe & P01      & $a, b$       \\
1351$+$318  & 20.2       & 1.14913 & 1.326   & $1.25         $ & E06b               & $-0.27\pm0.17$  & Zn & P99      & $d$          \\
1354$+$258  & 21.5       & 1.4205  & 2.006   & $0.61         $ & RT00               & $-1.58\pm0.16$  & Zn & K05      & $c$          \\
1451$+$123  & 19.9       & 2.255   & 3.246   & $1.20\pm0.08  $ & UVES$^{2}$         & $-1.08\pm0.23$  & Zn & D05      & $e$          \\
1622$+$238  & 20.4       & 0.6561  & 0.927   & $1.29         $ & SS92               & $-0.87\pm0.25$  & Fe & P03a     & $d$, 3C 336  \\
1629$+$120  & 19.7       & 0.9008  & 1.795   & $1.20\pm0.09  $ & R06                & $-0.18       $  & Zn & E06a     & $d$          \\
1727$+$5302 & 21.2       & 0.9448  & 1.444   & $2.832\pm0.070$ & R06                & $-0.52\pm0.2 $  & Zn & K04      & ---          \\
1727$+$5302 & 21.4       & 1.0312  & 1.444   & $0.922\pm0.057$ & R06                & $-1.39\pm0.2 $  & Zn & K04      & ---          \\
2206$-$1958 & 20.5       & 1.9205  & 2.56    & $1.99\pm0.05  $ & UVES$^{3,17,19}$   & $-0.37\pm0.07$  & Fe & P01      & ---          \\
2231$-$0015 & 20.5       & 2.066   & 3.015   & $1.80         $ & E06b               & $-0.86\pm0.10$  & Zn & D04      & $e$          \\
2343$+$125  & 20.3       & 2.4310  & 2.549   & $3.22\pm0.15  $ & UVES$^{6,12,17}$   & $-0.74\pm0.08$  & Zn & D04      & $a, e$       \\
2348$-$1444 & 20.6       & 2.279   & 2.94    & $0.32         $ & E06b               & $-1.84       $  & Fe & D06      & $d$          \\\hline\vspace{-4mm}
\end{tabular}
{\scriptsize
  Notes: $^a$\MOLH\ absorption detected; $^b$21-cm absorption
  detected; $^c$21-cm absorption not detected; $^d$radio-loud but not
  searched for 21-cm absorption; $^e$radio-quiet.
  $^{1}$60.A-9022 UVES commissioning; $^{2}$65.O-0063 Ledoux;
  $^{3}$65.O-0158 Pettini; $^{4}$65.P-0038 Srianand; $^{5}$66.A-0212
  Reimers; $^{6}$66.A-0594 Molaro; $^{7}$ 67.A-0022 D'Odorico;
  $^{8}$67.A-0146 Vladilo; $^{9}$68.A-0106 Petitjean; $^{10}$68.A-0361
  Lopez; $^{11}$68.A-0600 Ledoux; $^{12}$69.A-0204 D'Odorico;
  $^{13}$69.B-0108 Srianand; $^{14}$70.A-0017 Petitjean; $^{15}$70.B-0258
  Dessauges-Zavadsky; $^{16}$072.A-0100 Murphy; $^{17}$072.A-0346 Ledoux;
  $^{18}$073.B-0787 Dessauges-Zavadsky; $^{19}$074.A-0201 Srianand.\\
  References: SM79 \citep{SpinradH_79a}, B87 \citep{BouladeO_87a}, W93
  \citep{WolfeA_85a}, LB92 \citep{LanzettaK_92a}, MY92
  \citep{MeyerD_92a}, SS92 \citep{SteidelC_92a}, A94
  \citep{AldcroftT_94a}, M95 \citep{MeyerD_95a}, P97
  \citep{PettiniM_97a}, B98 \citep{BoisseP_98a}, L99
  \citep{LopezS_99a}, P99 \citep{PettiniM_99a}, PW99
  \citep{ProchaskaJ_99b}, C00 \citep{ChurchillC_00b}, D00
  \citep{delaVargaA_00a}, P00 \citep{PettiniM_00b}, RT00
  \citep{RaoS_00a}, P01 \citep{ProchaskaJ_01b}, P02
  \citep{PetitjeanP_02a}, L02a \citep{LedouxC_02b}, L02b
  \citep{LedouxC_02a}, C03a \citep{CenturionM_03a}, C03b
  \citep{ChurchillC_03b}, D03 \citep{Dessauges-ZavadskyM_03a}, L03
  \citep{LedouxC_03a}, LE03 \citep{LopezS_03a}, P03a
  \citep{ProchaskaJ_03b}, P03b \citep{ProchaskaJ_03d}, R03
  \citep{RyabinkovA_03a} [and references therein], D04
  \citep{Dessauges-ZavadskyM_04a}, J04 \citep{JunkkarinenV_04a}, K04
  \citep{KhareP_04a}, P04 \citep{PerouxC_04a}, A05
  \citep{AkermanC_05a}, K05 \citep{KulkarniV_05a}, L05
  \citep{LopezS_05a}, D06 \citep{Dessauges-ZavadskyM_06a}, E06a
  \citep{EllisonS_06a}, E06b (S.~E.~Ellison, private communication),
  P06a \citep{PerouxC_06a}, P06b \citep{ProchaskaJ_06a}, R06
  \citep{RaoS_06a}. `SDSS' or `UVES' indicates we have obtained \EW\
  from Sloan Digital Sky Survey spectra or from VLT/UVES spectra.
}
\end{minipage}
\end{center}
\end{table*}

We identified DLAs [$\NHI\ge 2\times10^{20}{\rm \,cm}^{-2}$] and
strong sub-DLAs [$\NHI\ge 3\times10^{19}{\rm \,cm}^{-2}$] with
existing metallicity measurements in the compilation of
\citet{KulkarniV_05a} and several other, typically more recent,
sources (as cited in Table \ref{dla}).  For these systems we searched
the literature for \EW\ measurements; Table \ref{dla} lists those
systems for which we found both [M/H] and \EW. Uncertainties for some
\EW\ values could not be found in the literature (see Section
\ref{ssec:corr}). For systems without a published \EW\ value we
searched for publicly available SDSS ($R\sim2000$) or Very Large
Telescope Ultraviolet and Visual Echelle Spectrograph (VLT/UVES; $R\ga
40000$) spectra from which an \EW\ measurement is easily made.  For
SDSS spectra, our method for obtaining \EW\ is described in detail in
\citet{BoucheN_06a}.  UVES spectra were obtained from the ESO data
archive and reduced using a version of the UVES pipeline which we
modified to improve the flux extraction and wavelength calibration.
The extracted echelle orders from all exposures were combined, using
{\sc uves popler}\footnote{Available at
  http://www.ast.cam.ac.uk/$\sim$mim/UVES\_popler.html}, with
inverse-variance weighting and a cosmic ray rejection algorithm, to
form a single spectrum with a dispersion of 2.5\,\kms\,pixel$^{-1}$.
\EW\ was then measured by direct summation of pixels across the Mg{\sc
  \,ii} $\lambda$2796 profile. The uncertainties in \EW\ quoted in
Table \ref{dla} for these UVES spectra are generally dominated by
uncertainties in the continuum level around the Mg{\sc \,ii}
$\lambda$2796 line. In some cases where \EW\ is so high that the
$\lambda$2796 profile is blended with that of the $\lambda$2803 line
(typically $\EW\ga 3.5$\,\AA), larger uncertainties are reported.

For most of the 49 absorbers in Table \ref{dla}, Zn-metallicities were
available in the literature. In 13 cases we have used the Fe or Cr
abundance in the absence of a Zn measurement. It is well known that
these elements are depleted onto dust grains
\citep[e.g.][]{PettiniM_97a} and so we corrected the Fe and Cr
abundances with the simple prescription of \citet{ProchaskaJ_03b}:
${\rm [M/H]} = {\rm [Fe/H]} + 0.4 = {\rm [Cr/H]} + 0.2$. In Section
\ref{ssec:corr} we test the possible impact this might have on our
results. Most of the metallicities in Table \ref{dla} have been
corrected to the solar scale of \citet{LoddersK_03a}. The values
quoted in the literature were used for Q\,0235$+$164, Q\,1622$+$238,
Q\,0458$-$020, Q\,1122$-$1649, Q\,1157$+$014 since we could not
identify which solar values the original authors used. Since the
corrections are typically $\la0.1{\rm \,dex}$, these 5 exceptions have
negligible impact on our main results.

\subsection{The $\bmath{\EW}$--[M/H] correlation}\label{ssec:corr}

\begin{figure*}
\hbox{
 \includegraphics[width=\columnwidth]{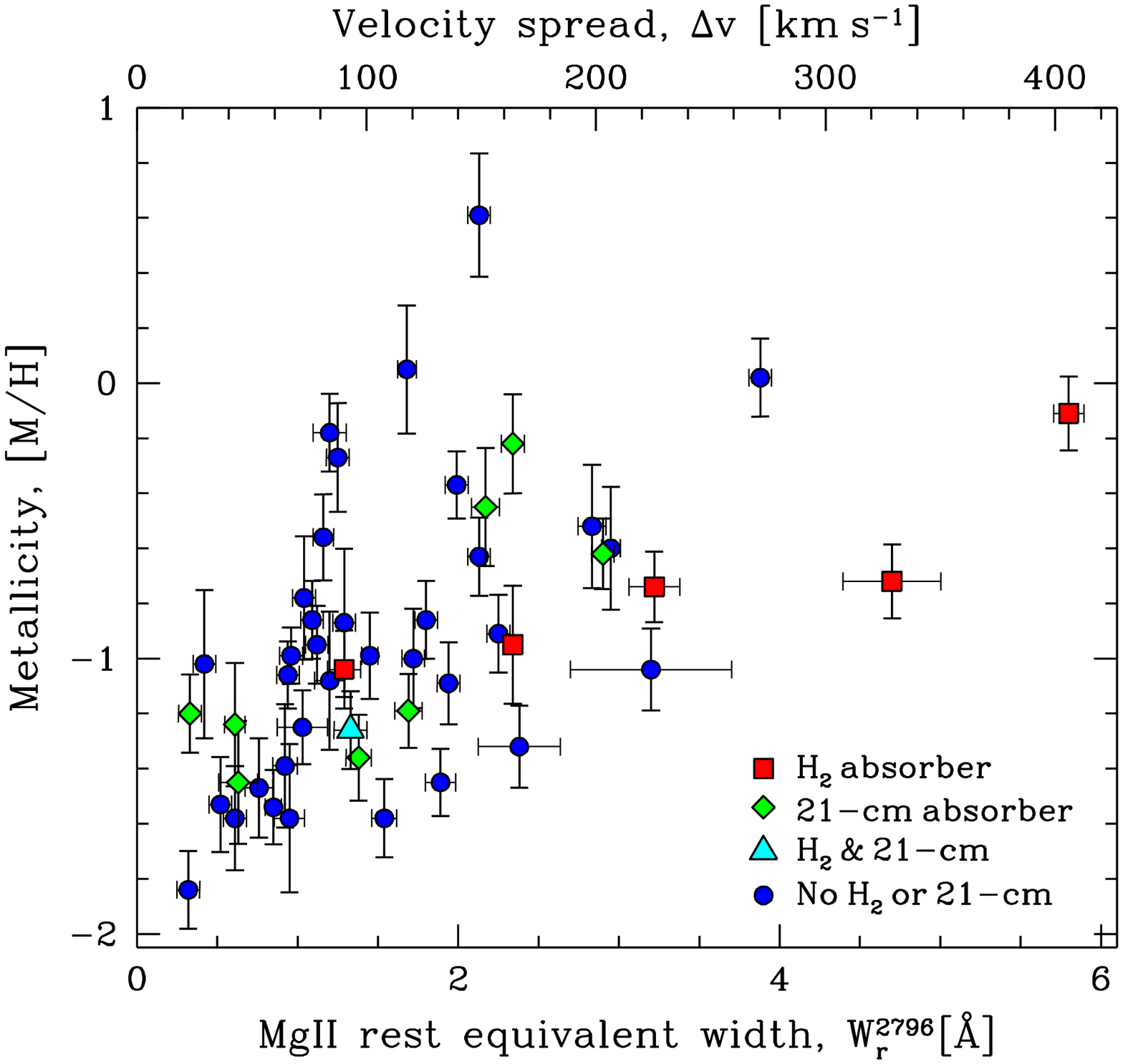}
 \hspace{3mm}
 \includegraphics[width=\columnwidth]{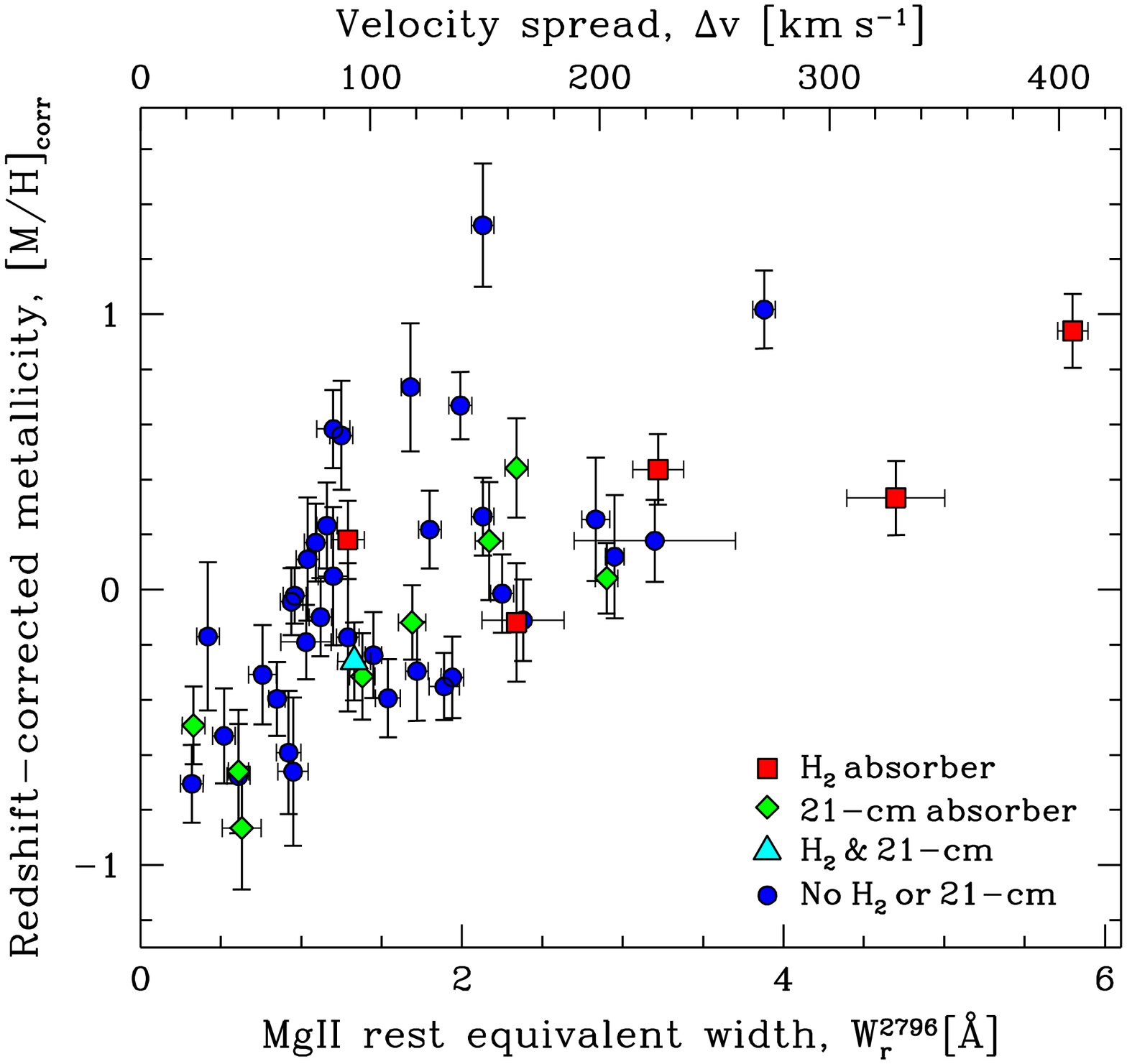}
}
\vspace{-2mm}
\caption{Correlation between absorber metallicity, [M/H], and Mg{\sc
    \,ii} rest-frame equivalent width, \EW. The upper velocity scale
  is the approximate velocity spread calculated simply from $\Delta
  v\,[\kms] \approx 70\,[\kms\,{\rm \AA}^{-1}] \times \EW\,[{\rm
    \,\AA}]$. {\it Left}: Original data from Table \ref{dla} with
  additional [M/H]- and \EW-uncertainties of $0.05{\rm \,dex}$ and
  $0.03$\,\AA\ added in quadrature (see text). The correlation is
  significant at the $4.2\,\sigma$ level. {\it Right}: The [M/H]
  values have been corrected for cosmological metallicity evolution
  according to equation (\ref{eq:evol}). This reduces the scatter,
  improving the correlation to the $4.7\,\sigma$ level.}
\label{M-W}
\end{figure*}

\begin{table*}
\begin{center}
\vspace{-2mm}
\begin{minipage}{0.99\textwidth}
  \caption{Statistics for different sub-samples. $N_{\rm a}$ is the
    sample size. $\left<\EW\right>$ and $\left<{\rm [M/H]}\right>$ are
    the median rest-frame Mg{\sc \,ii} equivalent width and the median
    metallicity, respectively. Columns 5--8 show the correlation
    statistics for the original data from Table \ref{dla} (with
    increased errors) while columns 9--12 show the same statistics
    after the metallicities are corrected for redshift evolution (see
    text). $P(\tau)$ is Kendall's $\tau$ two-sided probability that no
    correlation exists between [M/H] and \EW. We convert this to a
    significance, $S(\tau)$, assuming Gaussian statistics. In all
    cases a Spearman test gives a more significant correlation. $A$
    and $B$ are the intercept and slope, respectively, of the fit to
    the [M/H] versus $\Delta v$ data (see Section \ref{ssec:highz}).}
\label{stats}
\vspace{-2mm}
\begin{tabular}{l c c c c c c c c c c c c c c c c c c c}\hline
                                  &             &                    &                            & \multicolumn{4}{c}{Original data}                       & \multicolumn{4}{c}{Redshift-corrected metallicities}    \\
Sample                            & $N_{\rm a}$ & $\left<\EW\right>$ & $\left<{\rm [M/H]}\right>$ & $P(\tau)$  & $S(\tau)$ & $A$            & $B$           & $P(\tau)$  & $S(\tau)$ & $A$            & $B$           \\
                                  &             & [\AA]              & [dex]                      & [$10^{-3}$]& [$\sigma$]& [dex]          & [dex]         & [$10^{-3}$]& [$\sigma$]& [dex]          & [dex]         \\\hline
0: Fiducial                       & 49          & $1.38$             & $-0.99$                    & $0.026  $  & $4.2$     & $-4.29\pm0.38$ & $1.69\pm0.20$ & $0.0032 $  & $4.7$     & $-3.24\pm0.35$ & $1.62\pm0.18$ \\
1: $\left[{\rm Zn/H}\right]$ only & 36          & $1.33$             & $-0.97$                    & $0.48   $  & $3.5$     & $-4.28\pm0.46$ & $1.70\pm0.24$ & $0.038  $  & $4.1$     & $-3.14\pm0.40$ & $1.60\pm0.21$ \\
2: DLAs only                      & 40          & $1.42$             & $-1.02$                    & $0.028  $  & $4.2$     & $-3.85\pm0.28$ & $1.41\pm0.14$ & $0.0011 $  & $4.9$     & $-2.86\pm0.28$ & $1.38\pm0.14$ \\
3: $\EW > 0.7$\,\AA\              & 42          & $1.69$             & $-0.93$                    & $2.3    $  & $3.1$     & $-4.96\pm0.50$ & $1.99\pm0.25$ & $1.7    $  & $3.1$     & $-3.70\pm0.44$ & $1.83\pm0.22$ \\
4: $1+2+3$                        & 26          & $1.57$             & $-0.97$                    & $2.2    $  & $3.1$     & $-4.25\pm0.35$ & $1.59\pm0.17$ & $0.35   $  & $3.6$     & $-2.97\pm0.34$ & $1.44\pm0.16$ \\
5: $\EW < 1.4$\,\AA\              & 25          & $0.96$             & $-1.20$                    & $9.7    $  & $2.6$     & $-4.54\pm0.70$ & $1.91\pm0.39$ & $1.8    $  & $3.1$     & $-3.67\pm0.63$ & $1.94\pm0.35$ \\
6: $\EW > 1.4$\,\AA\              & 24          & $2.21$             & $-0.73$                    & $100    $  & $1.6$     & $-5.88\pm1.32$ & $2.33\pm0.58$ & $17     $  & $2.4$     & $-5.33\pm0.88$ & $2.50\pm0.39$ \\
7: $\zabs < 1.4$                  & 24          & $1.56$             & $-0.82$                    & $20     $  & $2.3$     & $-3.98\pm0.57$ & $1.64\pm0.30$ & $46     $  & $2.0$     & $-3.28\pm0.60$ & $1.67\pm0.31$ \\
8: $\zabs > 1.4$                  & 25          & $1.33$             & $-1.19$                    & $0.31   $  & $3.6$     & $-4.16\pm0.36$ & $1.52\pm0.19$ & $0.048  $  & $4.1$     & $-3.09\pm0.34$ & $1.53\pm0.17$ \\
9: No \MOLH\                      & 43          & $1.29$             & $-1.00$                    & $0.11   $  & $3.9$     & $-4.52\pm0.46$ & $1.83\pm0.24$ & $0.025  $  & $4.2$     & $-3.38\pm0.42$ & $1.71\pm0.22$ \\
10: No \MOLH\ or 21-cm            & 35          & $1.25$             & $-0.99$                    & $0.92   $  & $3.3$     & $-4.80\pm0.54$ & $1.97\pm0.29$ & $0.40   $  & $3.5$     & $-3.49\pm0.48$ & $1.78\pm0.25$ \\
11: \MOLH\ systems                &  6          & $2.78$             & $-0.84$                    & $15     $  & $2.4$     & $-4.06\pm0.63$ & $1.44\pm0.29$ & $91     $  & $1.7$     & $-3.26\pm0.75$ & $1.55\pm0.32$ \\
12: 21-cm systems                 &  9          & $1.38$             & $-1.20$                    & $95     $  & $1.7$     & $-3.65\pm0.68$ & $1.37\pm0.33$ & $12     $  & $2.5$     & $-2.74\pm0.63$ & $1.30\pm0.31$ \\\hline
\end{tabular}
\end{minipage}
\end{center}
\end{table*}

Figure \ref{M-W}(left) shows all the [M/H]--\EW\ pairs. Table
\ref{dla} gives the formal statistical errors on both quantities as
reported in the literature. In cases where no error-bars could be
found in the literature, uncertainties of 0.10\,dex and 0.05\,\AA\
were assumed as these typified the rest of the sample. For
Fig.~\ref{M-W}(left) we added 0.05\,dex and 0.03\,\AA\ in quadrature
to the initial [M/H] and \EW\ uncertainties to take into account
likely continuum fitting errors which are not included by some authors
or, indeed, by our own \EW\ measurements in SDSS spectra. We regard
the points in Table \ref{dla} with these increased errors as our
fiducial sample. Kendall's $\tau$ for the [M/H]--\EW\ correlation
evident in Fig.~\ref{M-W} has a probability $P(\tau) = 2.6\times
10^{-5}$ of being due to chance alone; assuming Gaussian statistics,
the correlation is significant at $>\!4.2$-$\sigma$. The correlation
is robust to the removal of distinct sub-samples, some of which are
discussed below, as shown in Table \ref{stats}.

The absorbers cover quite a large redshift range, $0.2<\zabs<2.6$, so
evolution in DLA metallicities may have somewhat washed out the
correlation we observe. For example, if absorbers at lower \zabs\ are,
on average, more metal-rich than at higher \zabs, then this would
cause additional dispersion in [M/H] at all values of \EW. This would
at least decrease the statistical significance of the observed
correlation and, depending on the redshift distribution of the sample,
may somewhat alter its slope. To explore whether this is indeed the
case, we make a simple correction to the metallicities in Table
\ref{dla} according to
\begin{equation}\label{eq:evol}
{\rm [M/H]}_{\rm corr} = 0.52 + {\rm [M/H]} + 0.27\,\zabs\,.
\end{equation}
The numerical coefficients are the $y$-intercept and slope from an
unweighted least squares fit to [M/H] versus $\zabs$ using the
fiducial sample. This has the effect of correcting each metallicity by
the expected mean value at the absorber's redshift. The slope and
intercept are consistent with those derived from larger samples of
DLAs covering similar redshift ranges
\citep[e.g.][]{ProchaskaJ_03b,KulkarniV_05a}. Fig.~\ref{M-W}(right)
shows the redshift-corrected metallicities versus \EW\ and Table
\ref{stats} gives the redshift-corrected statistics: the general
scatter around the [M/H]--\EW\ correlation clearly decreases upon
correction; the table shows that the significance of the [M/H]--\EW\
correlation increases in almost all sub-samples once the metallicities
are corrected for redshift evolution.

Visually, the correlation in Fig.~\ref{M-W}(left) appears tighter for
the lower-\EW\ half of the sample compared to that for the upper half.
That is, the variance in [M/H] about the general [M/H]--\EW\ trend
appears smaller at low-\EW.  However, an F-test reveals only marginal
evidence for this: a fit of [M/H] versus \EW\ is performed [see
equation (\ref{eq:fit}); Section \ref{ssec:highz}], the sample is
split into two sub-samples at $\EW=1.4$\,\AA\ and the variance in
[M/H] about the fit is calculated for each sub-sample. The ratio of
the variances about the fit is $1.83$, but this should occur 15\,per
cent of the time by chance alone given the sizes of the sub-samples.
Using the redshift-corrected metallicities [Fig.~\ref{M-W}(right)],
the ratio of the variances reduces to $1.64$, which has an associated
probability of 24\,per cent. The marginally increased scatter at high
\EW\ does reduce the significance of the correlation in the high-\EW\
sub-sample compared to the low-\EW\ sub-sample (see Table
\ref{stats}); one possibility is that Fig.~\ref{M-W}, particularly the
redshift-corrected version in Fig.~\ref{M-W}(right), defines a lower
bound `envelope' to the metallicity at a given \EW\ instead of a
normal correlation. This should of course be combined with an upper
bound envelope at high metallicities since arbitrarily high values are
unphysical. A larger sample with robust control over metallicity
uncertainties would be required to identify these features and to
distinguish between a true correlation and a lower bound envelope.

We have also split the sample into two redshift bins about the
fiducial sample's median, $\zabs=1.4$. Table \ref{stats} shows that
the [M/H]--\EW\ correlation is still well-defined in the high-$\zabs$
sub-sample but it is somewhat less statistically significant at
low-$\zabs$. However, the fits to both sub-samples, as described in
Section \ref{ssec:highz}, yield consistent results.

Three main objections might be raised to some of the data we use in
Table \ref{dla}: (i) Some of the metallicities are not derived from Zn
and so might be susceptible to dust-depletion effects, even though we
have made a crude correction to any metallicities derived using the
heavily depleted Fe and Cr ions; (ii) Absorbers with low \EW\ might
contain a significant number of velocity components which are not
completely saturated. While \EW\ still provides a measure of velocity
spread, $\Delta v$, in these systems, \EW\ will also depend somewhat
on the optical depths of the unsaturated components.  Therefore, as
the MgII velocity components weaken at lower [M/H], one might expect a
correlation between [M/H] and \EW\ which has little to do with the
correlation found at higher \EW. One might consider that only systems
with $\EW > 0.7$\,\AA\ should be used since the profile is even more
likely to be completely saturated; (iii) Finally, the reliability of
sub-DLA metallicities is not as well established in the literature as
that for DLAs. It is possible that we do not fully understand the
ionization corrections required in some cases or, indeed, the
magnitude of the systematics involved in deriving \NHI\ in sub-DLAs.
For example, Voigt profile \NHI\ estimates closer to the flat part of
the curve of growth naturally have larger random uncertainties and may
become increasingly, systematically influenced by weaker Ly-$\alpha$
forest lines.  However, Table \ref{stats} shows that the [M/H]--\EW\
correlation is fairly robust against removal of non-Zn metallicities,
systems with $\EW < 0.7$\,\AA\ or sub-DLAs.  Removing all three cases
together also leaves a $>3\,\sigma$ correlation ($>3.5\,\sigma$ after
correcting the metallicities for redshift evolution).

Note that our sample contains only one system with ${\rm [M/H]} \la
-1.7$.  This is partially a selection effect since the Zn{\sc \,ii}
lines become undetectably weak at low metallicities. However, very few
DLAs and sub-DLAs are known with lower metallicities at redshifts $0.2
< \zabs < 2.6$ and comparison with other more complete samples
\citep[e.g.][]{KulkarniV_05a,AkermanC_05a} shows that our sample is
fairly representative of the overall DLA metallicity distribution.
Finally, we have included 3 systems in Table \ref{dla} which are
within 5000\,\kms\ of the emission redshift of the background QSO.
These systems do not show any signs of being associated with the QSO
central engine and so are included here. Removing them from the sample
makes a negligible difference to our results.

\section{Discussion}\label{sec:disc}

The [M/H]--\EW\ correlation found in Fig.~\ref{M-W} is a direct
demonstration that the metallicity of strong Mg{\sc \,ii} absorbers is
closely related to their kinematics (but not necessarily the
kinematics of the host galaxy; see below). Since the Mg{\sc \,ii}
$\lambda$2796 transition is so strong, most velocity components in
absorbers with $\EW > 0.3$\,\AA\ are saturated and so \EW\ measures
the total velocity spread among the components, $\Delta v$.
Equivalently, it is a measure of the number of velocity components
across the profile \citep[e.g.][]{PetitjeanP_90a}.  Thus, the
[M/H]--\EW\ correlation demonstrates that the mechanism responsible
for producing individual Mg{\sc \,ii} velocity components and for
dispersing them over velocity ranges $\Delta v \sim 20$--$1000$\,\kms\
is related to -- and possibly also determines -- the metallicity of
the absorber.

The results in Fig.~\ref{M-W} bear on the apparent discrepancy between
the conclusions drawn by \citet{NestorD_03a}, whose results suggest a
[M/H]--\EW\ correlation, and \citet{YorkD_06a} whose analysis suggests
an anti-correlation. Our observed [M/H]--\EW\ correlation supports the
former and seems inconsistent with the latter. One possible
explanation may be that \citeauthor{YorkD_06a} assume a constant
dust-to-gas ratio as a function of \NHI, i.e.~$E(B-V)\propto\NHI$.
They base this assumption on their observation that the Mg{\sc \,ii}
absorbers they study seem to redden the background QSOs with, on
average, an SMC-like dust extinction law. Even if this proves to be
true for strong Mg{\sc \,ii} absorbers, this relation has certainly
not been demonstrated in DLAs or sub-DLAs.  Indeed, from the detection
of SMC-like dust-reddening in $>300$ SDSS DLAs at $\zabs>2.2$, Murphy
et al.~(2007, in preparation) find that the mean
$E(B-V)\approx7\times10^{-3}$ appears not to change over the range
$20.3\le\log[\NHI/{\rm cm}^{-2}]\le21.7$. One caveat may be important
here: possible magnitude bias. \citeauthor{NestorD_03a} assumed that
\NHI\ is not a strong function of \EW\ based on the observations of
\citet{RaoS_06a}.  The data in Table \ref{dla} also reveal no apparent
trend in \NHI\ with \EW. However, \citeauthor{RaoS_06a} selected
relatively bright background QSOs which could provide reasonable
signal-to-noise ratios in ultraviolet (UV) {\it Hubble Space
  Telescope} (HST) spectra.  \citeauthor{YorkD_06a} suggest that dusty
absorbers are therefore unlikely to exist in \citeauthor{RaoS_06a}'s
sample and that this dust-bias (relative to deeper surveys like SDSS)
may invalidate \citeauthor{NestorD_03a}'s assumption. From this point
of view, it may not be surprising that our results agree with the
conclusion of \citeauthor{NestorD_03a} since most, but not all, of the
absorbers in our sample occult relatively bright QSOs.

\subsection{Comparison with higher redshift DLAs}\label{ssec:highz}

At higher redshifts ($1.6 < \zabs < 3.0$), \citet{WolfeA_98a}
tentatively noted a correlation between $\Delta v$ and metallicity
from a sample of 17 DLAs. \citet{PerouxC_03b} found a similarly
tentative [Zn/H]--$\Delta v$ correlation from a larger sample of DLAs
and sub-DLAs over the redshift range $1.4<\zabs<4.5$.  However,
\citet[][hereafter \citetalias{LedouxC_06a}]{LedouxC_06a} very
recently found a relatively tight and clear [M/H]--$\Delta v$
correlation from a sample of UVES spectra containing 70 DLAs and
sub-DLAs over the range $1.7 < \zabs < 4.3$. They defined $\Delta v$
from different transitions of several low ionization species (O{\sc
  \,i}, Si{\sc \,ii}, Fe{\sc \,ii}, Cr{\sc \,ii}, S{\sc \,ii}) with
moderate optical depths such that the strongest velocity component
absorbed between 10 and 60\,per cent of the continuum. Since many of
the DLAs in the \citetalias{LedouxC_06a} sample are at $z_{\rm abs} >
2.6$ where the Mg{\sc \,ii} lines are not observable with optical
spectroscopy, a complete comparison between the
\citetalias{LedouxC_06a} sample and ours is difficult. Nevertheless,
it is clear that, since \EW\ is a measure of $\Delta v$, the
[M/H]--\EW\ correlation in Fig.~\ref{M-W} persists at higher
redshifts.

\citetalias{LedouxC_06a} fit their [M/H]--$\Delta v$ data using the
least squares bisector method of \citet{IsobeT_90a}, finding the
best-fit relationship to be ${\rm [M/H]} = (1.55\pm0.12) \log\Delta
v\,[\kms]-(4.33\pm0.23)$. For comparison, we performed fits using the
same bisector technique after converting our \EW\ values to $\Delta v$
according to the simple relation $\Delta v\,[\kms] \approx
70\,[\kms\,{\rm \AA}^{-1}] \times \EW\,[{\rm \,\AA}]$. The
proportionality constant is derived from a fit of \EW\ versus $\Delta
v$, with the inverse of the variances on \EW\ used as weights, for the
18 absorbers which appear in both our sample and that of
\citetalias{LedouxC_06a}. The fiducial sample is best-fitted by the
relation
\begin{equation}\label{eq:fit}
  {\rm [M/H]} = (1.69\pm0.20) \log\Delta v\,[\kms]-(4.29\pm0.38)\,,
\end{equation}
which is consistent with the relation found by \citetalias{LedouxC_06a}.

Table \ref{stats} shows the intercept $A$ and slope $B$ of the
[M/H]--$\Delta v$ relationship for the different sub-samples. In
particular, we find no evidence for significant differences in $A$ or
$B$ in the low- and high-$\zabs$ sub-samples. This is similar to the
results of \citetalias{LedouxC_06a} who find the slope and intercept
to be statistically consistent in two redshift bins split at
$\zabs=2.43$.  We also note that the median \EW\ and [M/H] of our
low-$\zabs$ sub-sample are marginally higher than at high-$\zabs$:
$\left<\EW\right>=1.56$\,\AA\ and $\left<{\rm [M/H]}\right>=-0.82$ for
the former compared with $\left<\EW\right>=1.33$\,\AA\ and $\left<{\rm
    [M/H]}\right>=-1.19$ for the latter. Again,
\citetalias{LedouxC_06a} find similar trends.

\subsection{Mass--metallicity correlation or anti-correlation?}

The [M/H]--\EW\ (or [M/H]--$\Delta v$) correlation in DLAs and
sub-DLAs provides an important link between the kinematics {\it of the
  absorber} and the metal-enrichment history {\it of the absorber}.
However, one would like to go a step further and link the metallicity
with the mass of the halo in which the absorber resides. This step is
potentially confusing. For example, many authors have, in the past,
tacitly assumed that the absorption-line kinematics are a reliable
tracer of the host-galaxy kinematics. That is, $\Delta v$ is assumed
to positively correlate with the circular velocity of the host-galaxy.
By presuming this detail of the absorber--galaxy connection -- which
should instead be observationally determined -- one would conclude
that \EW\ (and $\Delta v$) is correlated with the galaxy mass.
\citetalias{LedouxC_06a} make this latter assumption and use it to
constrain the implied mass--metallicity relationship. Indeed, they
find this to be consistent with the luminosity--metallicity
relationship derived from local galaxies.

However, the assumption that galaxy or halo mass is correlated with
\EW\ is challenged by new observations.  \citet[][hereafter
\citetalias{BoucheN_06a}]{BoucheN_06a} identified 1806 Mg{\sc \,ii}
systems with $\EW \ge 0.3$\,\AA\ and $\sim 250000$ luminous red
galaxies (LRGs) within projected co-moving distances of $13h^{-1}{\rm
  \,Mpc}$ of the absorbers. The ratio of the absorber--LRG
cross-correlation to the LRG--LRG auto-correlation provides a measure
of the bias-ratio between absorbers and LRGs. In hierarchical
structure formation scenarios the bias scales with halo mass, so the
clustering of LRGs around the Mg{\sc \,ii} absorbers provides a
measure of the absorber halo mass. With this approach,
\citetalias{BoucheN_06a} found that halo mass and \EW\ are {\it
  anti-correlated}. They also found that by combining the previously
observed luminosity- and \EW-dependence of the absorber cross-section,
or by combining the cross-section's luminosity-dependence with the
observed incidence probability of strong Mg{\sc \,ii} absorbers, ${\rm
  d}^2N/{\rm d}\zabs{\rm d}\EW$, one finds that luminosity (or mass)
and \EW\ should be anti-correlated. \citet{ProchterG_06a} reached a
similar conclusion by interpreting the observed redshift evolution of
the number density of strong Mg{\sc \,ii} systems with simple
cross-section arguments.

Of course, the main difference between the \citetalias{BoucheN_06a}
absorber sample and the one studied here is that
\citetalias{BoucheN_06a} used strong Mg{\sc \,ii} absorbers while we
study DLAs and strong sub-DLAs. It is possible that a mass--\EW\
correlation does exist for DLAs if the sharp drop in halo-mass
observed by \citetalias{BoucheN_06a} for Mg{\sc \,ii} systems with
$\EW > 1.0$\,\AA\ were dominated by systems with neutral hydrogen
column densities below the lower limit imposed here,
i.e.~$\NHI=3\times10^{19}{\rm \,cm}^{-2}$. However, this seems very
unlikely given that the fraction of strong Mg{\sc \,ii} systems which
are DLAs [i.e.~have $\NHI=2\times10^{20}{\rm \,cm}^{-2}$] {\it
  increases} from $\sim\!15$ to $\sim\!65$\,per cent over the range
$\EW=0.6$--$3.3$\,\AA\ \citep{RaoS_06a}. It is therefore important to
confirm or refute the \citetalias{BoucheN_06a} results, which derive
from SDSS Data Release 3. Bouch{\' e} et al.~(2007, in preparation) will
present an analysis based on a significantly increased sample of
absorbers from Data Release 5.

\subsection{Are strong Mg{\sc \,ii} absorbers caused by outflows?}

\begin{figure*}
\centerline{\includegraphics[width=\textwidth]{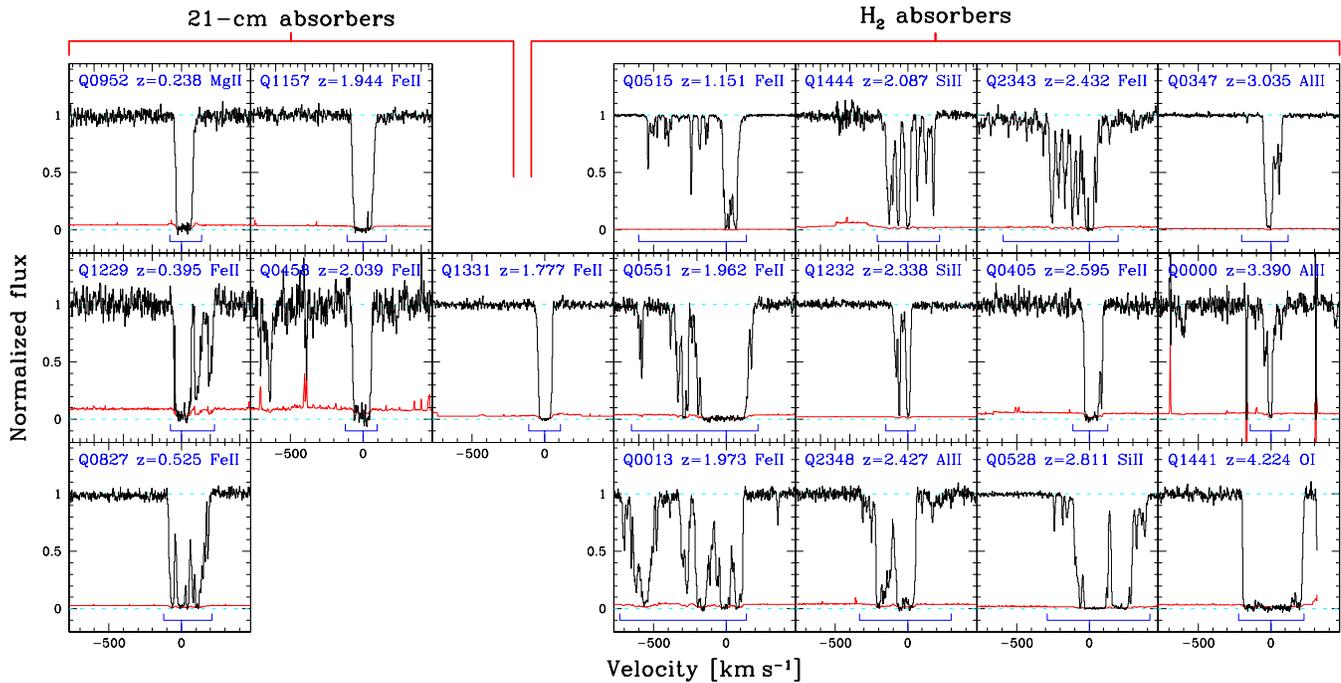}}
\vspace{-3mm}
\caption{Montage of the velocity structures of strong transitions in
  the 13 known \MOLH-bearing DLAs/sub-DLAs in the literature (6 are in
  our sample) and the 21-cm absorbers in our sample. The transition
  used to illustrate the extent and distribution of absorption
  components is marked on each panel. In most cases the Fe{\sc \,ii}
  $\lambda$2382 line is used. Fe{\sc \,ii} $\lambda$2600 is used for
  the absorber towards Q\,0458$-$020. When Fe{\sc \,ii} $\lambda$2382
  was not available (particularly at high-\zabs), O{\sc \,i}
  $\lambda$1302, Si{\sc \,ii} $\lambda$1526, Al{\sc \,ii}
  $\lambda$1670 or Mg{\sc \,ii} $\lambda$2796 is used. All systems are
  shown on a common velocity scale registered to the redshift of the
  strongest \MOLH\ or 21-cm component in each system. The black
  histogram is the data normalized by a local continuum fit while the
  grey/red line near zero flux shows the normalized $1\,\sigma$ error
  array. The approximate extent of the transitions is marked by the
  bracket below the zero level. Note the similarity in the profiles
  for many of the \MOLH-bearing systems. These broad and complex
  structures may be characteristic of outflows. The 21-cm profiles
  generally do not show such outflow signatures, but the sample is too
  small for strong conclusions to be made.}
\label{profiles}
\end{figure*}

\citetalias{BoucheN_06a} interpret their observed mass--\EW\
anti-correlation as evidence that outflows become the dominant
mechanism for producing Mg{\sc \,ii} absorbers as \EW\ increases
beyond $\sim1.0$\,\AA. They reason that outflows are more easily
ejected from lower mass star-forming galaxies and so, when viewed
along a random QSO sight-line, are more likely to produce more Mg{\sc
  \,ii} velocity components over a larger velocity range. In this
outflow picture, \EW\ and the number of Mg{\sc \,ii} velocity
components may be inversely related to the mass and could depend more
strongly on the recent star-formation history of the host-galaxy.
\citetalias{LedouxC_06a} mention that some fraction of DLAs could be
caused by outflows, but they again attribute the outflow velocity
spread to be directly related to the depth of the host galaxy's
potential well.

An initial objection to the outflow model for DLAs and strong Mg{\sc
  \,ii} systems might be that such hot, highly ionized material may
not produce velocity components of cold gas which can be traced with
low-ionization species such as Mg{\sc \,ii}. However, the outflows
from local starburst galaxies are well-known to contain clouds of cold
gas and dust
\citep{LehnertM_96a,MartinC_99a,HeckmanT_00a,RupkeD_05a,MartinC_06a}.
Dense, molecular gas is also known to be entrained in smaller scale
outflows \citep[e.g.][]{NakaiN_87a} and even the ionization cones of
nearby active galactic nuclei which show starburst activity
\citep[e.g.][]{IrwinJ_92a,CurranS_99a}.  Moreover, \citet{NormanC_96a}
identified Mg{\sc \,ii} absorption in spectra of two QSO sight-lines
with impact parameters $25\,\hkpc$ and $55\,\hkpc$ near the galaxy NGC
520, a local starburst with super-winds, a disturbed morphology and
filamentary H$\alpha$ emission along its minor axis. Thus, it seems
clear that outflows from star-forming galaxies do contain cold gas.

If indeed many high-\EW\ Mg{\sc \,ii} absorbers are caused by outflows
and these outflows do indeed entrain cold clouds (or cold gas clumps
cool out of the hotter outflow medium), do we observe tracers of the
cold, dusty gas, such as molecular hydrogen? \MOLH\ is identified via
UV absorption in the Lyman and Werner bands and is known to arise in
very cold, dusty velocity components \citep[e.g.][]{LedouxC_03a}. It
is therefore notable that 3 of the 5 highest-\EW\ systems in
Fig.~\ref{M-W} are known to contain \MOLH.  Indeed, the two
highest-\EW\ systems, those towards Q\,0013-0029 and Q\,0551-366, both
contain \MOLH\ \citep{LedouxC_02a,PetitjeanP_02a} and have \EW\ values
at the extreme of the \EW\ distribution obtained from large SDSS
Mg{\sc \,ii} surveys \citep[e.g.][]{NestorD_03a,ProchterG_06a}.  The
median \EW\ for the \MOLH\ absorbers in our sample is $2.8$\,\AA\
compared with $1.4$\,\AA\ for the sample as a whole (Table
\ref{stats}). Even with only six \MOLH\ systems, a Kolmogorov-Smirnov
test gives a probability of just 9\,per cent that the equivalent
widths of the \MOLH\ absorbers are drawn from the same parent
population as the rest of the sample.

In Fig.~\ref{profiles} we compare the velocity structures of all the
13 known H$_2$-bearing DLAs and sub-DLAs. It is striking how similar
the velocity structures are for the systems towards Q\,0515$-$4414,
Q\,0551$-$366 and Q\,0013$-$0029: each is spread over $\Delta v \ga
700$\,\kms\ and comprises at least three groups of stronger velocity
components separated by few (if any) weaker components. The systems
towards Q\,1444$+$014 and Q\,2343$+$125 have very similar profiles and
these are not dissimilar to the above three.  Also, the systems towards
Q\,2348$-$011 and Q\,0528$-$250 both have two broad, saturated regions
separated by a narrow region of low optical depth. Such structures and
symmetries have been discussed by \citet{BondN_01a} and
\citet{EllisonS_01e} as being the characteristic signs of absorption
in galactic outflows. It is also conceivable that these profiles are
produced in galaxy mergers \citep[but see discussion in][]{BondN_01a}.

Although no strong conclusions can be reached due to the sample size,
Fig.~\ref{profiles} nevertheless suggests that approximately half of
the known H$_2$-bearing systems may arise in outflows. If indeed cold
gas and dust is entrained in outflows and the gas is well-shielded
then H$_2$ should form on the surface of dust-grains. Even if this
happens in very few clouds, the large number of clouds across these
profiles significantly increases the probability of observing one or
more H$_2$-bearing clouds towards any given background QSO.  Hence,
even if these outflow-like absorption profiles are themselves rare,
H$_2$ may be found within them in a large fraction of cases.

Fig.~\ref{profiles} also shows the velocity structures of the 21-cm
absorbers in our sample for which UVES spectra were available. These
velocity structures are typical of those in the general population,
seemingly consistent with the fact that the 21-cm absorbers are
generally indistinct from most other systems in the [M/H]--\EW\ plane
(Fig.~\ref{M-W}). Like H$_2$ absorption, 21-cm absorption should arise
in cold clouds, but there is no evidence in Fig.~\ref{profiles} to
suggest that 21-cm absorbers also arise in the complicated and broad
outflow-like profiles like the H$_2$ systems. However, the 21-cm
sample is very small and is dominated by low-$\zabs$ systems; outflows
are likely to be less common below $\zabs\sim 1$ due to the general
decline in the global star-formation rate density \cite[see,
e.g.,][]{BondN_01a}.

\citetalias{LedouxC_06a} note that, under the assumption of a
mass-metallicity relationship, the absorbers with the highest
equivalent widths and metallicities will be most easily detected in
direct imaging observations. This prediction is also natural in the
outflow picture where \EW\ and the number of Mg{\sc \,ii} velocity
components are related to both the mass and the recent star-formation
history of the absorber host-galaxy. While one expects the highest
\EW\ and highest [M/H] systems to be hosted by lower mass galaxies,
they should be actively star-forming and show strong H$\alpha$
emission. If indeed H$_2$-bearing systems more often probe
star-forming galaxies with outflows, they may be more reliable tracers
of the cosmological evolution of metallicity than other DLAs, many of
which might not be closely linked to star-formation sites/processes in
high-$z$ galaxies.  We have already advanced this possibility in
\citet{CurranS_04c} and \citet{MurphyM_04b} based on a comparison of
H$_2$ absorber metallicities with those of the general DLA population.
The H$_2$ absorbers showed a faster, more well-defined increase in
[M/H] with decreasing redshift, although the sample was quite small. A
much larger sample of H$_2$-bearing systems is required to compute the
\NHI-weighted metallicity evolution in a way comparable with general
DLA samples.

\section{Conclusions}\label{sec:conc}

For 49 DLAs and sub-DLAs with published metallicities we have measured
or gathered from the literature their Mg{\sc \,ii} rest-frame
equivalent widths to search for an empirical relationship between the
kinematic spread, $\Delta v$, in the absorbing metal-line velocity
components and the total absorber metallicity. The vast majority of
DLAs and strong sub-DLAs have $\EW>0.3$\,\AA\ which means that most
Mg{\sc \,ii} velocity components across the absorption profile are
saturated. Thus, \EW\ provides a simple measure of $\Delta v$ which
can be derived even from relatively low-resolution QSO spectra.

We find a correlation between [M/H] and \EW\ at the significance level
of $4.2\,\sigma$ [Fig.~\ref{M-W}(left)] which increases to
$4.7\,\sigma$ when the [M/H] values are corrected for mild
cosmological evolution over the sample's redshift range of
$0.2<\zabs<2.6$ [Fig.~\ref{M-W}(right)]. Even for our most
conservative sub-sample, where we include only 26 DLAs (no sub-DLAs)
with Zn-metallicities and $\EW>0.7$\,\AA, we still find a correlation
at $>3\,\sigma$ ($>3.5\,\sigma$ after redshift-correction of [M/H]).
After converting the \EW\ values to $\Delta v$, the slope of our
[M/H]--$\log\Delta v$ relationship is consistent with that reported
recently by \citet{LedouxC_06a} for generally higher-\zabs\ DLAs and
sub-DLAs, suggesting that the [M/H]--\EW\ correlation persists up to
$\zabs=4.7$. Although we find a lower statistical significance for the
correlation below $\zabs=1.4$, the fitted slope and intercept of the
[M/H]--$\log\Delta v$ data is similar for the $\zabs<1.4$ and
$\zabs>1.4$ sub-samples. Finally, we find marginal ($\sim1.4\,\sigma$)
evidence for increased variance in metallicity above $\EW=1.4$\,\AA.
However, the evidence weakens once the [M/H] values are corrected for
evolution; other non-intrinsic sources of metallicity scatter may need
to be addressed before such an effect can be verified.

If one assumes that the metal-line absorption kinematics trace the
host-galaxy kinematics -- that $\Delta v$ is proportional to the
host-galaxy circular velocity -- then it is natural to translate the
observed [M/H]--\EW\ correlation into a correlation between [M/H] and
galaxy mass. Indeed, \citet{LedouxC_06a} make this assumption and find
that the resulting [M/H]--mass correlation is consistent with the
[M/H]--luminosity relationship derived from local galaxies. However,
by studying the clustering of galaxies around strong Mg{\sc \,ii}
absorbers in the SDSS, \citet{BoucheN_06a} find that mass is
anti-correlated with \EW. This result, together with the [M/H]--\EW\
relationship observed here, implies that the absorber metallicity is
anti-correlated with halo or galaxy mass. It is therefore important to
confirm or refute the results of \citeauthor{BoucheN_06a} and to
address the assumption of \citet{LedouxC_06a} with observations.

Finally, we note that about half of the known \MOLH\ absorbers have
very broad velocity structures which show distinct distributions of
velocity components (Fig.~\ref{profiles}). The patterns, groups and
symmetries evident in these cases are consistent with an outflow
origin for the bulk of the components. Since outflows can entrain
cold, dusty gas and since they should provide a large number of
velocity components, the probability of finding one or more
\MOLH-bearing components in any given outflow-driven absorber might be
quite high, even though \MOLH-bearing components are themselves quite
rare. This suggests the targeting of high-\EW\ absorbers to identify
more \MOLH-bearing systems. Given the [M/H]--\EW\ correlation we
observe, this may be similar to the metallicity-selection approach of
\citet{PetitjeanP_06a}. However, it has the advantage that one can
determine \EW\ with only moderate-resolution spectra -- or infer it
from the equivalent width of bluer transitions when $\zabs\ga2.5$ --
which need not be blue- or UV-sensitive (e.g.~SDSS spectra) . One can
then follow-up the highest \EW\ systems with high-resolution, UV- or
blue-sensitive spectra to search for \MOLH\ absorption.

\section*{Acknowledgments}

We thank N.~Bouch\'{e} for many discussions and critical comments
which significantly improved the paper and S.~Ellison for providing
her \EW\ measurements. We also thank H.-W.~Chen for comments on an
early draft. MTM thanks PPARC for an Advanced Fellowship.  Some of
this research was based on observations made with ESO Telescopes at
the Paranal Observatories under programme IDs listed in Table
\ref{dla}. This research has made use of the NASA/IPAC Extragalactic
Database (NED) which is operated by the Jet Propulsion Laboratory,
California Institute of Technology, under contract with the National
Aeronautics and Space Administration.  This research has also made use
of NASA's Astrophysics Data System Bibliographic Services.


\bsp_small

\label{lastpage}

\end{document}